\documentclass[9pt,twocolumn,twoside]{pnas-new}
\usepackage{sistyle}
\usepackage{overpic}
\usepackage{pdfpages}

\templatetype{pnasresearcharticle} 

\title{Jump Rope Vortex in Liquid Metal Convection}

\author[a,b,1]{Tobias Vogt}
\author[a,1,2]{Susanne Horn} 
\author[a]{Alexander M. Grannan}
\author[a]{Jonathan M. Aurnou}

\affil[a]{Department of Earth, Planetary, and Space Sciences, University of California, Los Angeles, CA 90095, USA}
\affil[b]{Institute of Fluid Dynamics, Helmholtz-Zentrum Dresden-Rossendorf, 01328 Dresden, Germany}

\leadauthor{Vogt}

\significancestatement{The large-scale circulation (LSC) is the key dynamical feature of turbulent thermal convection. It is the underlying structure that shapes the appearance of geo- and astrophysical systems, such as the solar granulation or cloud streets, and the cornerstone of theoretical models.
Our laboratory-numerical experiments reveal for the first time that the LSC can perform a fully three-dimensional motion resembling a twirling jump rope. The discovery of this novel LSC mode implies that the currently accepted paradigm of a quasi-planar oscillating LSC needs to be revisited. Moreover, it provides an important link between studies in confined geometries used in experiments and simulations and the virtually unconfined fluid layers in natural settings where an agglomeration of LSCs forms larger patterns.}

\authorcontributions{T.V., A.M.G. and J.M.A. designed the experiments. T.V. and A.M.G. performed the experiments. S.H. designed and performed the numerical simulations. All authors analysed and interpreted the results and wrote the paper.}
\authordeclaration{The authors declare no conflict of interest.}
\equalauthors{\textsuperscript{1}T.V. and S.H. contributed equally to this work.}
\correspondingauthor{\textsuperscript{2}To whom correspondence should be addressed. E-mail: t.vogt@hzdr.de}


\begin{abstract}
Understanding large scale circulations (LSCs) in turbulent convective systems is important for the study of stars, planets and in many industrial applications. The canonical model of the LSC is quasi-planar with additional horizontal sloshing and torsional modes
[Brown E, Ahlers G (2009) \textit{J. Fluid Mech.} 638:383--400; Funfschilling D, Ahlers G (2004) \textit{Phys. Rev. Lett.} 92(19):194502; Xi HD et al. (2009) \textit{Phys. Rev. Lett.} 102(4):044503; Zhou Q et al. (2009) \textit{J. Fluid Mech.} 630:367--390].
Using liquid gallium as the working fluid, we show via coupled laboratory-numerical experiments that the LSC in a tank with aspect ratios greater than unity takes instead the form of a ``jump rope vortex'', a strongly three-dimensional mode that periodically orbits around the tank following a motion much like a jump rope on a playground. 
Further experiments show that this jump rope flow also exists in more viscous fluids such as water, albeit with a far smaller signal. 
Thus, this new jump rope mode is an essential component of the turbulent convection that underlies our observations of natural systems. 
\end{abstract}

\dates{This manuscript was compiled on \today}
\doi{\url{www.pnas.org/cgi/doi/10.1073/pnas.XXXXXXXXXX}}

\begin{document}

\maketitle
\thispagestyle{firststyle}
\ifthenelse{\boolean{shortarticle}}{\ifthenelse{\boolean{singlecolumn}}{\abscontentformatted}{\abscontent}}{}

\dropcap{I}n fully turbulent convecting systems, convective energy coalesces into coherent, large scale flows. These fluid motions manifest as superstructures in most geo- and astrophysical systems, where they form characteristic patterns such as the granulation on the Sun's surface or cloud streets in the atmosphere \cite{vonHardenberg2008, Pandey2018, Stevens2018}. The underlying building block of these superstructures is a singular large-scale circulation (LSC), 
first observed over 35 years ago in the laboratory experiments of Krishnamurti and Howard \cite{Krishnamurti1981}.
The LSC, also called the ``wind of turbulence'' \cite{Grossmann2000}, is the largest overturning structure in a given fluid layer. Despite the similar shape, the turbulent LSC is distinct from the laminar convection rolls that develop at convective onset \cite{Xi2004, Villermaux1995, Funfschilling2004, Zhou2009}. An agglomeration of LSCs can then act to form a superstructure in spatially extended fluid layers \cite{Pandey2018, Stevens2018, vonHardenberg2008, Sakievich2016, Bailon2010}.

In all natural and experimental turbulent convecting fluid systems, LSCs have been found to exist \cite{Cioni1997, Tsuji2005, Schumacher2016, Funfschilling2004, Horn2013, Breuer2009}.
They have been described as having a quasi-two-dimensional, vertical planar structure whose flow follows a roughly circular or elliptical path \citep{Villermaux1995, Funfschilling2004, Brown2009, Zhou2009, Xi2009}. Within contained systems, the quasi-planar flow rises vertically on one side of the container and descends vertically on the antipode. All LSCs have a regular low-frequency oscillation that is the dominant spectral signature of the flow \citep{Heslot1987}. In recent years this low-frequency oscillation has been characterised as arising from a misalignment of the vertical symmetry plane of the LSC, resulting in a horizontal sloshing and torsioning of the LSC flow \citep{Funfschilling2004, Sun2005, Brown2009, Xi2009, Zhou2009, Stevens2011}.  In a decoupled view of the motions, the torsioning of the plane resembles a sheet of paper being alternately twisted and counter-twisted around the central axis of the container, and the horizontal sloshing mode resembles a purely vertical sheet moving side-to-side through the fluid, perpendicular to the LSC plane. It is argued that these two oscillatory modes are coupled such that they exist simultaneously in any given system \citep{Xi2009, Brown2009, Zhou2009}.

Often considered the most important dynamical attribute of thermal convection, the LSC flow shears the upper and lower boundary of the fluid layer \cite{Schumacher2016}. This shearing promotes the emission of new thermal plumes, which, in turn, helps to drive the mean wind. Many theoretical approaches rely on the LSC concept to predict key output quantities, such as the net convective heat and momentum transport \citep{Grossmann2000, Ahlers2009, Shishkina2015}. Hence, having an accurate descriptions of the LSC structure  and the complete range of possible dynamics is mandatory. 

\begin{figure}[t!]
\centering
\includegraphics[width=.9\linewidth]{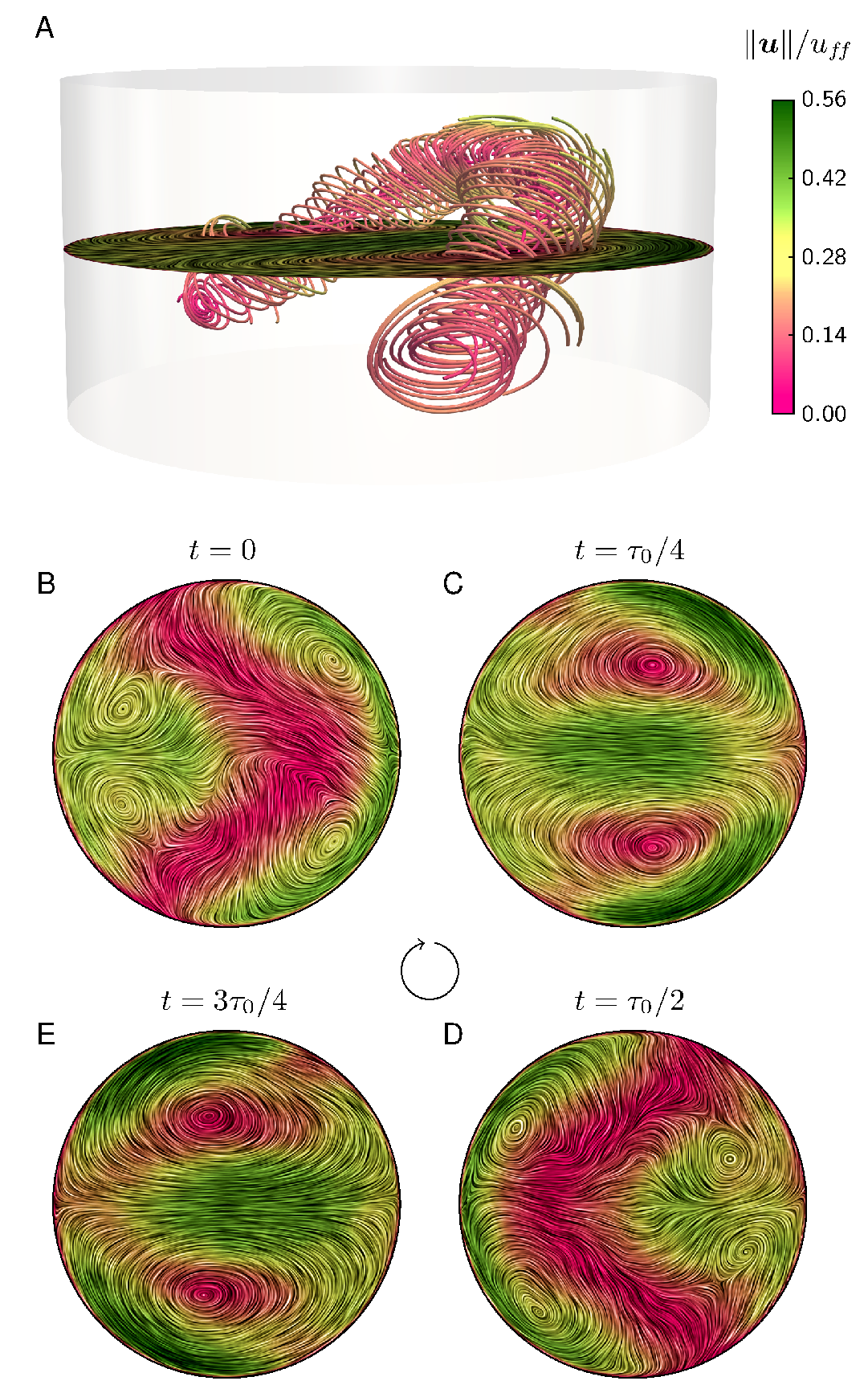}
\caption{The jump rope vortex. Panel (A) shows a conditionally-averaged three-dimensional (3D) visualisation of the streamlines at $t=\tau_0/4$ from DNS made with $Pr = 0.025$, $Ra = 1.12 \times 10^6$ in a cylindrical tank with a diameter-to-height aspect ratio $\Gamma = D/H = 2$. The streamlines surround the jump rope vortex core, with streamline color denoting local velocity magnitude. In addition, in the midplane slice,  colour contours denote velocity magnitude and velocity streamlines are visualised by the line-integral convolution (LIC) method. 
The jump rope cycle is shown in panels (B)~$\rightarrow$ (C) $\rightarrow$ (D) $\rightarrow$ (E) $\rightarrow$ (B). Shown are cross-sections at half height, for the same conditionally-averaged phases as the coloured sidewall profiles in Fig.~\ref{fig:temperature}\textit{B}. The vortex core of the LSC has minimal velocity (pink). In (B), $t=\tau_0$ and (D), $t=\tau_0/2$, the LSC is confined to the midplane. In (C), $t=\tau_0/4$, and (E), $t=3\tau_0/4$, the LSC has moved out of the midplane. The highest velocities (green) in (C/E) also show the clear splitting of the up and downwelling flows. (See Supplementary Movies S2 and S3 and Supplementary Fig. S2 for 3D renderings of panels (B)--(D).)}
  \label{fig:velocity}
\end{figure}

\section*{Laboratory-Numerical Convection Experiments}
We study Rayleigh-B\'enard convection (RBC), in which a fluid layer is heated from below and cooled from above. The system is defined by three non-dimensional control parameters: the Rayleigh number $Ra = \alpha g \Delta H^3 / (\kappa \nu)$, which describes the strength of the buoyancy forcing relative to thermal and viscous dissipative effects; the Prandtl number $Pr = \nu/\kappa$, which discribes the thermophysical fluid properties; and the container's aspect ratio $\Gamma = D/H$. Here, $\alpha$ is the isobaric thermal expansion coefficient, $\nu$ is the kinematic viscosity, $\kappa$ is the thermal diffusivity, $g$ us the gravitational acceleration, $\Delta$ is the temperature drop across the fluid layer, $D$ is the diameter and $H$ is the height of convection tank.

We have carried out combined laboratory-numerical RBC experiments using gallium, a low $Pr \simeq 0.027$ liquid metal, as the working fluid.  
The fluid is contained within a $\Gamma = 2$ cylinder, the largest container in which a single LSC will form and in which the highest convective velocities are attained \cite{Bailon2010, Sakievich2016, Stevens2018, Pandey2018}. This experimental design, which elaborates on the liquid mercury investigation of Tsuji et al. \cite{Tsuji2005} and departs from canonical $Pr \sim 1$, $\Gamma \sim 1$ studies, allows a sole LSC to develop in a maximally unconfined, strongly turbulent environment. The large amplitude velocity and temperature signatures in this system enable us to detect and quantify novel modes of the LSC that may not have been easily observed in the canonical set-up. In fact, we find a fundamentally new mode of large-scale turbulent convection with a three-dimensional oscillation that deviates from the quasi-planar description of LSC motion (Fig. \ref{fig:velocity}). Instead of sloshing or twisting side-to-side motions, our results show a flow that circulates around a crescent-shaped vortex, which in turn orbits the tank in the direction opposite the fluid velocity.  As seen in Supplementary Movie S3, this vortex looks like a twirling jump rope.

\begin{figure}[t]
\centering
  \begin{overpic}[width=\linewidth]{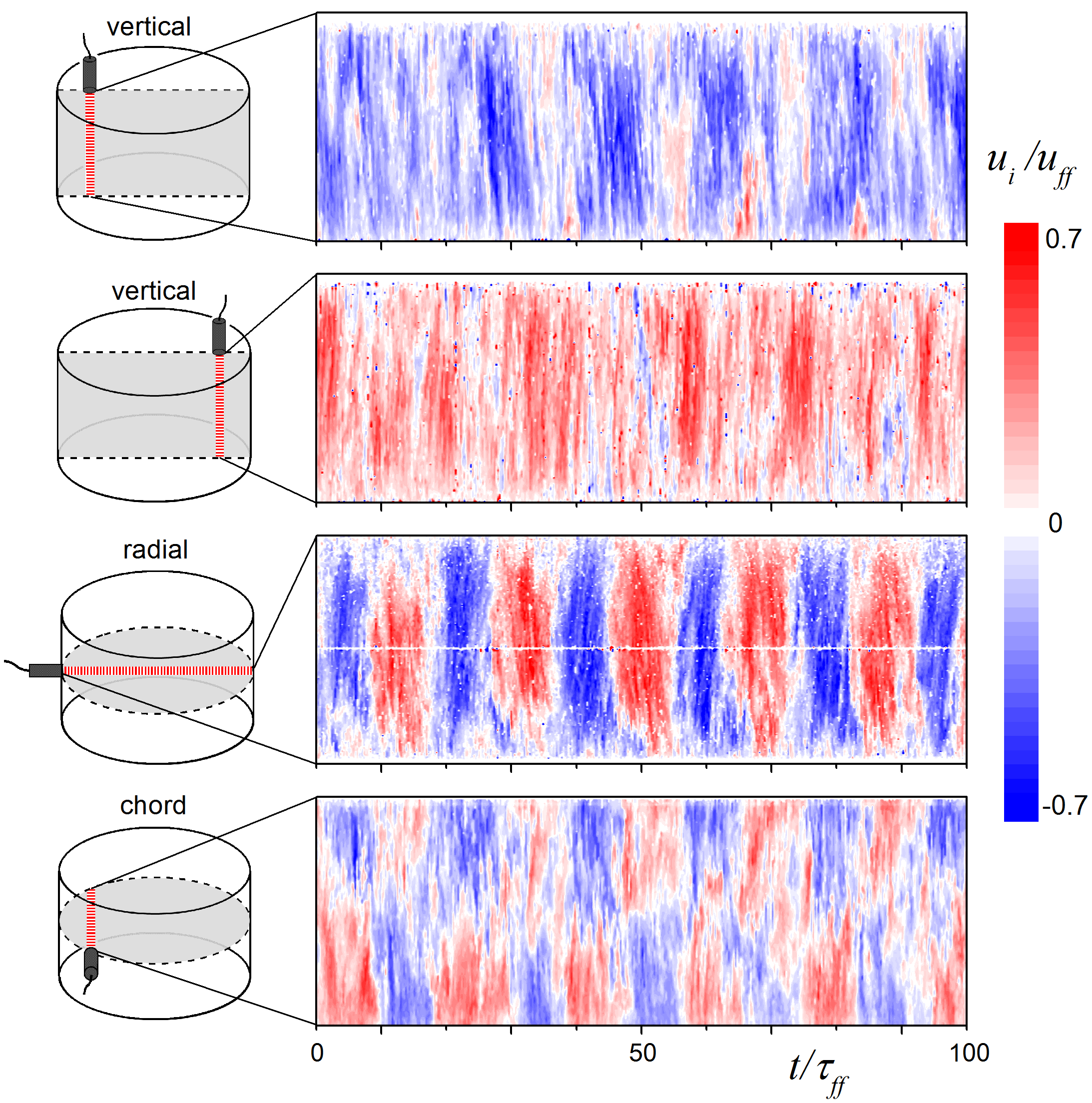}
   \put(1,96.9){\small \textsf{A}}
   \put(1,72.7){\small \textsf{B}}
   \put(1,49.3){\small \textsf{C}}
   \put(1,26.3){\small \textsf{D}}
   \end{overpic}
  \caption{Spatio-temporal evolution of laboratory convection velocities. Simultaneous ultrasonic Doppler measurements for the case at $Pr = 0.027$, $Ra~=~1.03 \times 10^6$ and $\Gamma = 2$. The particular measuring lines are indicated as a dashed red line in the schematics left of each data panel. Negative (positive) velocities represent flow towards (away from) the transducer. The measurements are non-dimensionalised using the free-fall velocity $u_{f\!f}=\sqrt{\alpha g \Delta H}$ and the free-fall time $\tau_{f\!f}= H/u_{f\!f}$. The ordinate corresponds to the measuring depth along the tank height (A, B), diameter (C), and chord (D), respectively. The measurements  in (A--C) lie in the symmetry plane of the LSC; the chord probe measurements in (D) lie perpendicular to the LSC symmetry plane. The axial velocity in (A) and (B) show a mean up- and downflow, respectively, and relatively weak periodic fluctuations. While the mean velocity is zero in (C) and (D), strong oscillations are observed that span the tank.  The flow along the chord in (D) shows a periodic double cell structure whereby the oscillation is in phase to (A). The strong periodic oscillation in (C) and (D) cannot be explained via the current LSC paradigm. (The white horizontal stripe in (C) is due to the standing echo from the \SI{1}{mm} diameter center point thermistor.)}
  \label{fig:Doppler}
\end{figure}
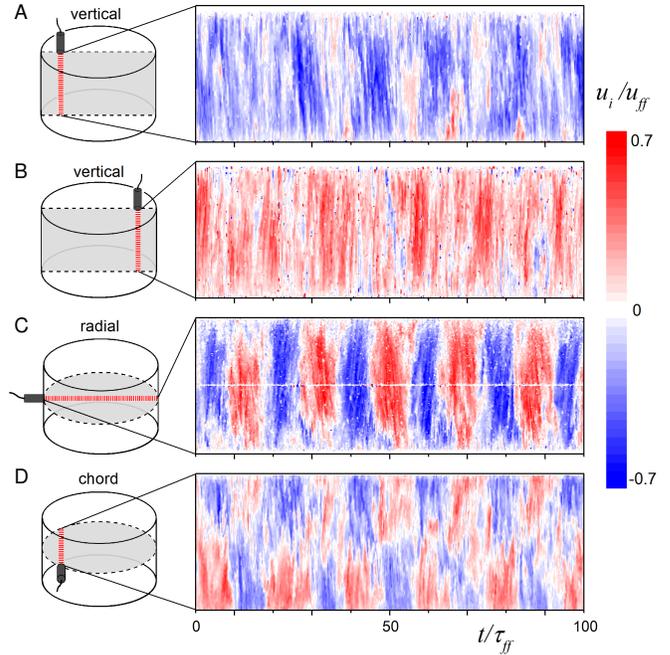
 
Our $\Gamma = 2$ liquid gallium laboratory experiments are performed on the RoMag device (see Methods) and span a Rayleigh number range of $7.1 \times 10^4 \leq Ra \leq 5.1 \times 10^6$. Ultrasonic Doppler velocimetry \citep{Vogt2018} is utilised to measure the instantaneous flow distribution along four different measuring lines (Fig.~\ref{fig:Doppler}). Two ultrasonic transducers are attached antipodaly to the upper end block at cylindrical radius $r/R = 2/3$ to measure the vertical velocity field (Fig.~\ref{fig:Doppler}\textit{A, B}). Another two transducers are fixed to the sidewall horizontally at height $z/H = 1/2$ to measure midplane velocities along the diameter (Fig.~\ref{fig:Doppler}\textit{C}) and along a chord (Fig.~\ref{fig:Doppler}\textit{D}). The transducers are all oriented to align approximately with the symmetry plane of the large-scale flow, except for the chord probe that is perpendicular to it. Additionally, 29 thermistors are used to measure the experimental temperature field, including the central temperature of the bulk fluid, the vertical temperature difference across the fluid layer, and along one-third of the mid-plane sidewall.  With this set-up, we diagnose the 3D dynamics of the liquid metal LSC.

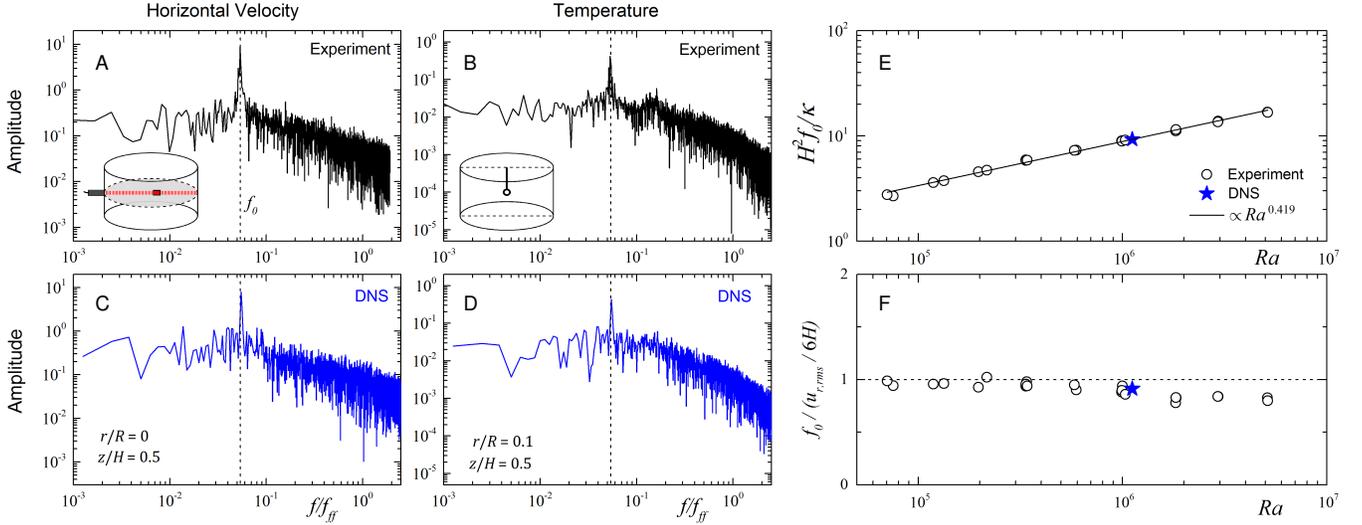
\begin{figure*}
  \centering
  \begin{overpic}[width=17.8cm]{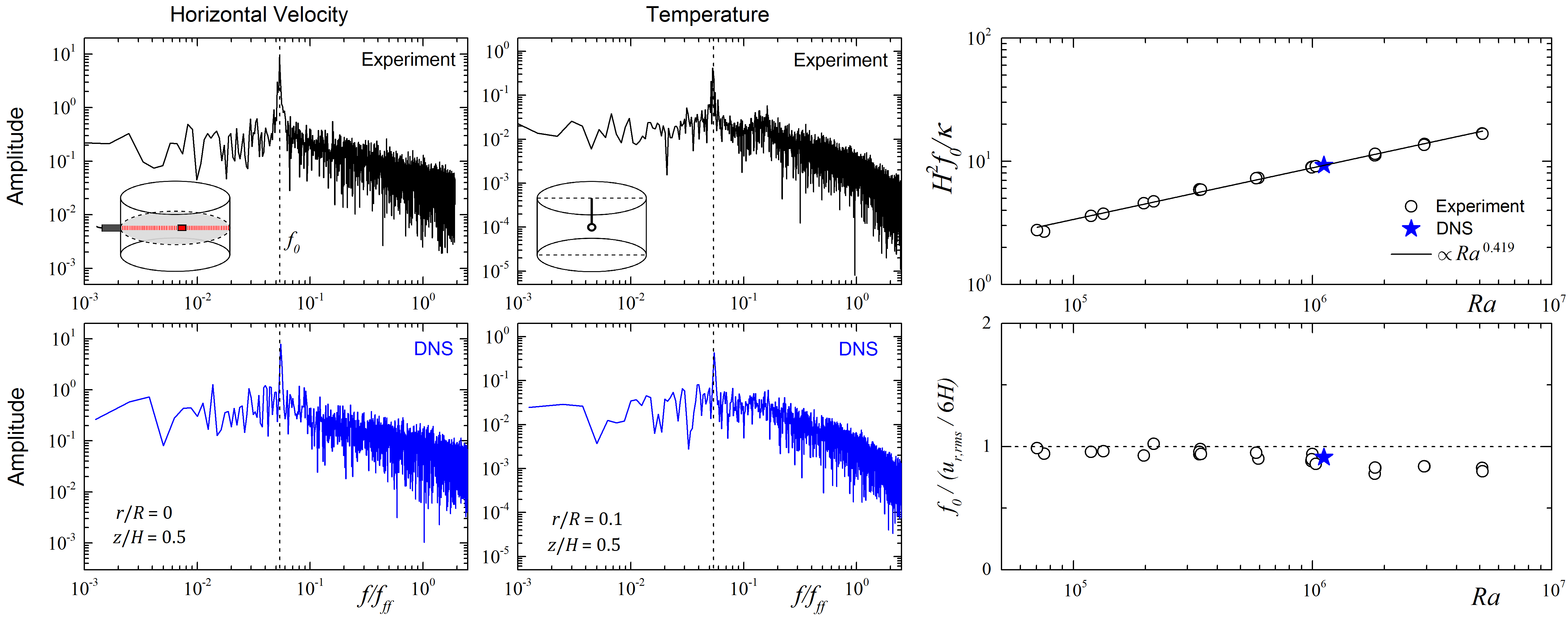}
   \put(7,34){\small \textsf{A}}
   \put(34.5,34){\small \textsf{B}}
   \put(65.5,34){\small \textsf{E}} 
    \put(7,16){\small \textsf{C}}
   \put(34.5,16){\small \textsf{D}}
   \put(65.5,16){\small \textsf{F}} 
   \end{overpic}
  \caption{Characteristic frequency and its scaling. Measured velocity and temperature frequency spectra in (A) and (B), respectively, in laboratory data at $Ra~=~1.03 \times 10^6$ and in (C) and (D) from DNS with $Ra = 1.12 \times 10^6$. All spectra are calculated using data obtained near the center of the fluid domain.
The four dashed lines in (A)--(D) indicate $f_0$ taken from (A). 
(E) shows the dominant frequency as a function of $Ra$, normalised by the thermal diffusion time, $H^2/\kappa$, for experiments (open circles) and DNS (blue star). The best fit of the experimental data yields a $Ra^{0.419}$ scaling. (F) shows the frequency normalised by the estimated LSC turnover time, $6H/u_{r, rms}$, where we use the maximum possible path length of $2H + 2D = 6H$.}
  \label{fig:freq}
\end{figure*}

All of our ultrasonic Doppler results exhibit a distinctive velocity pattern, visualised in Fig.~\ref{fig:Doppler} for $Ra = 1.03 \times 10^6$. The vertical velocities in Fig.~\ref{fig:Doppler}\textit{A, B} show flow near the axial plane of the LSC, with \textit{A} representing the cold downwelling flow (blue) and \textit{B} representing the warm upwelling motions (red) of the LSC. In addition, our measurements reveal both high and low frequency oscillations within the vertical velocity fields. The higher frequency oscillations correspond to small-scale plumes, whereas the lower frequency signals correspond to the fundamental oscillatory modes of the LSC. These vertical velocity measurements are in agreement with the quasi-planar model of the LSC. However, we find that the low frequency oscillation is strongest in the horizontal direction, aligned along the LSC's horizontal mid-plane (Fig.~\ref{fig:Doppler}\textit{C}). Further, the mid-plane chord probe measurements (Fig.~\ref{fig:Doppler}\textit{D}) show that the horizontal velocity switches sign across the midpoint of the chord, which lies in the symmetry plane of the LSC. These data indicate that the fluid is periodically diverging from the axial LSC plane and then converging back towards it. The measured velocities approach the free-fall velocity \citep{Glazier1999} $u_{f\!f} = \sqrt{\alpha g \Delta H}$. These velocities are well within our measurement capabilities and are thus detected as robust features of the flow. Significantly, these diverging-converging chord-probe flows indicate the presence of a strongly 3D flow pattern that is inconsistent with the previously proposed horizontal sloshing and torsional modes \citep{Brown2009, Xi2009}, and, thus, requires a novel physical explanation. 

To diagnose the modes of behaviour of the inertial LSC flow within the opaque liquid metal, we carried out high resolution direct numerical simulations (DNS), using the fourth order finite volume code \textsc{goldfish} to provide detailed information on the spatially and temporally fully-resolved three-dimensional temperature and velocity fields; see Methods. The main DNS employs parameter values of $\Gamma = 2$; $Ra = 1.12 \times 10^6$ and $Pr = 0.025$, and is run for 1000 free fall time units after reaching statistical equilibrium. 

We compare the outputs from the DNS and the laboratory experiments in Fig.~\ref{fig:freq}. Measured near the central point of the fluid bulk, the horizontal velocity (Fig.~\ref{fig:freq}\textit{A, C}) and temperature (Fig.~\ref{fig:freq}\textit{B, D}) spectra of the two systems quantitatively agree.

In fact, the peak oscillation frequency $f_0$ (dotted vertical line) is identical in all four spectra. This quantitative agreement demonstrates that our DNS captures the essential behaviours of the laboratory experiments, and is well-suited as a diagnostic tool to interpret the flows existing in the opaque liquid metal. In addition, the agreement shows that the idealised boundary conditions available in the DNS are sufficiently replicated in the laboratory experiments.

As a further verification of our coupled laboratory-numerical system, we compare the $f_0$ values (Fig.~\ref{fig:freq}\textit{A}) from our entire suite of experiments to the Grossmann--Lohse model for the bulk properties of LSC-dominated convective flows \citep{Grossmann2000, Stevens2013}. The value of $f_0$ is connected to the characteristic velocity of the LSC (Fig.~\ref{fig:freq}\textit{E, F}) and, thus, to the momentum transport \citep{Grossmann2002}. This transport is expressed by the Reynolds number $Re$, i.e. $f_0 H^2/\kappa = c Re Pr$ with a constant $c$ determined by the geometry, where $\kappa$ is the fluid's thermal diffusivity. 
The best fit to the Grossmann-Lohse model \cite{Grossmann2002, Stevens2013} predicts an effective scaling in our parameter range of $Re \propto  Ra^{0.435 \pm 0.002}$, which corresponds here to $f_0 \propto Ra^{0.422 \pm 0.002}$. By linear regression of our data set, we find that $f_0 \propto Ra^{0.419 \pm 0.006}$, or, equivalently, $Re \propto Ra^{0.433 \pm 0.006 }$ (Fig.~\ref{fig:freq}\textit{E}). Thus our DNS and experimental results are in agreement with time-averaged, kinematic models of LSC-dominated convective flows\citep{Grossmann2000}, but broaden the understanding of the time-varying LSC dynamics. 

Similar to previous studies, we characterise the LSC via temperature measurements acquired along the midplane circumference of the cylinder \citep{Brown2005, Stevens2011, Zhou2009, Xi2009}. These sidewall temperature measurements provide information about the large-scale convective flows in the interior of the convection cell. Because of the high thermal diffusivity of liquid metals, the large-scale temperature signal is exceptionally clear as the small-scale temperatures are damped by diffusion. In addition, the high velocities in inertial liquid metal convective flows strongly advect the large-scale temperature field, producing midplane temperatures that almost reach the imposed maximal temperatures that exist on the top and bottom boundaries. Thus, temperature signals provide a strong and clear window into the LSC dynamics in liquid metal flows.  

\begin{figure}[bh!]
  \centering
  \includegraphics[width=.9\linewidth]{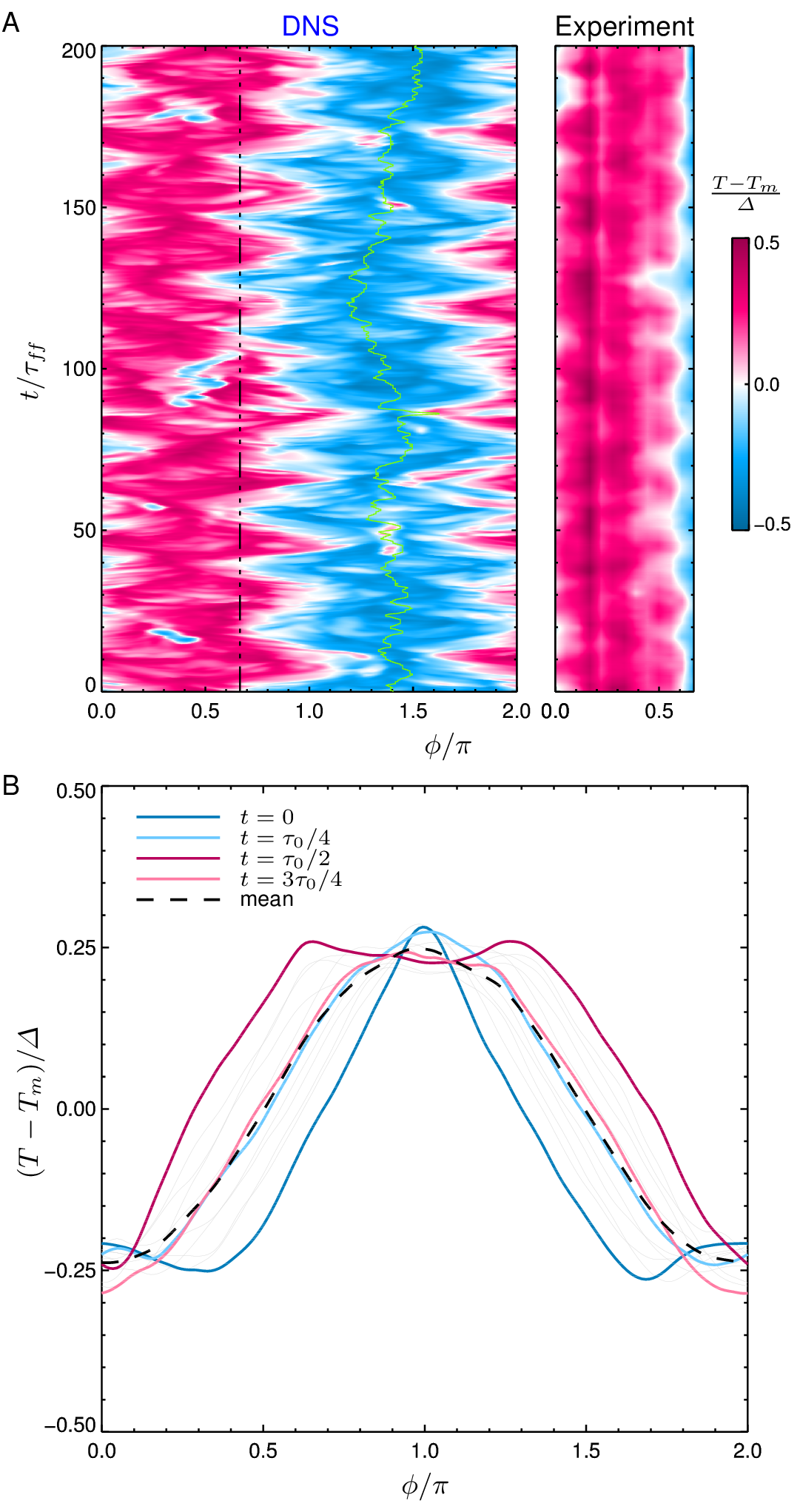}
  \caption{Midplane temperature signal on the sidewall. (A) Numerically and experimentally obtained temperature $(T-T_m)/\Delta$ at half height shown for 200 free-fall time units $t/\tau_{f\!f}$. The black dash-dotted line indicates the azimuthal $\phi$-width covered by the thermocouple array in the laboratory experiment. 
The instantaneous position of the LSC $\xi_{LSC}$ is demarcated by the green line, which exhibits relatively small meanderings.
In contrast, the jump roping of the LSC causes the thermal pattern to fluctuate strongly, with warmer fluid (then colder fluid) occupying between $1/3$ and $2/3$ of the circumference, creating what looks like the baffles of an accordion. (B) Conditionally averaged DNS profiles of the temperature sidewall for the 12 phases of the jump rope oscillation. They reveal the splitting of the cold LSC downflow at $t=0$ (dark blue) and hot LSC upflow at $t=\tau_0/2$ (dark magenta) by clear double minima and maxima, respectively. The disparity of the profiles at $t=\tau_0/4$ (light blue), $t=3\tau_0/4$ (light magenta) suggests that the motion is three-dimensional. The grey lines are the profiles for the remaining 8 phases of the oscillation. The total mean (black dashed line) results in a simple cosine function.}
  \label{fig:temperature}
\end{figure}

\section*{Jump Rope Vortex Cycle}
Fig.~\ref{fig:temperature}\textit{A} shows the midplane sidewall temperatures in the DNS (left) at $Ra = 1.12 \times 10^6$ and in the laboratory experiment (right) at $Ra = 1.03 \times 10^6$. The distinct blue and pink regions (cold and hot, respectively) reveal strong temperature alternations within the fluid, which we call an accordion pattern.  
At $t/\tau_{f\!f} = 0$, the cold fluid covers roughly two-thirds of the sidewall circumference; by $t/\tau_{f\!f} \simeq 10$, the cold fluid covers only one-third of the circumference and then subsequently expands in azimuth again. 
Averaging the sidewall temperatures over the entire circumference, $T_{a\!v\!g}  = \langle T(r=R, z=H/2) \rangle_\phi$, consequentially yields quasi-periodic oscillations in $T_{a\!v\!g}$; see Methods. Unlike cases dominated by torsion and sloshing modes \citep{Funfschilling2004, Brown2009, Xi2009, Zhou2009}, our sidewall measurements have the same frequency $f_0$ as found in the central measurements shown in Fig.~\ref{fig:freq}\textit{A}--\textit{D} (see Supplementary Movies S2 and S3). The accordion thermal pattern in the sidewall temperature pattern (Fig.~\ref{fig:temperature}\textit{A} and Supplementary Fig.~S1) and the corresponding oscillation in $T_{a\!v\!g}$ (Fig.~\ref{fig:conditional}) provide the simplest means by which to identify the jump rope mode in convection data. 

To fully diagnose the 3D characteristics of the LSC in the DNS, we first determined the position of the LSC within the fluid domain for each snapshot. Extending the single-sinusoidal-fitting method of Cioni et al. \cite{Cioni1997} and Brown et al. \cite{Brown2005}, each snapshot's midplane sidewall temperature distribution is fit to the function
\begin{equation*}
T_{fit}(\phi)  = a \cos(\phi - \xi_{LSC}) + b \cos(2 (\phi - \xi_{LSC})) + T_{a\!v\!g} \, ,
\end{equation*}
where $\xi_{LSC}$ denotes the symmetry plane of the LSC (i.e., the green line in Fig.~\ref{fig:temperature}\textit{A} and also in Supplementary Movie S1).  The values of $a$ and $b$ give the relative amplitudes of the hot and cold sidewall signals. 
We then decompose the oscillation period $\tau_0 = 1/f_0$ into 12 characteristic phases based on the rms value of the sidewall mid-plane temperature (see Methods). Using these 12 phases, $10\,000$ temporally equidistant snapshots are conditionally-averaged over a total time span of $t/\tau_{f\!f} = 1000$. In this operation, we first orient the solutions so that the LSC symmetry plane remains fixed in azimuth and then average the temperature and velocity fields at the same temporal phase values, as shown in Supplementary Movie S2. 

Our 12 conditionally-averaged midplane, sidewall temperature profiles are shown in Fig.~\ref{fig:temperature}\textit{B}. In agreement with previous quasi-planar LSC studies, the average of all phases yields a cosine function (dashed black line). However, our conditional averaging also reveals a more complex thermal structure with three extrema in the $t=0$ and $t = \tau_0/2$ profiles, with either a double-maxima-single-minimum or double-minima-single-maximum structure. Thus our conditional averaging extracts new information about the presence of additional complexity within the LSC dynamics. Further, a pure cosine function is not present at $t=\tau_0/4$ and $t=3\tau_0/4$, requiring the existence of spatiotemporal complexity in the LSC flow field.  

The convective flow can only be understood by considering the LSC as a fully 3D vortical structure, whose vortex core traces a path similar to that of a jump rope that precesses around the tank in the direction opposite that of the LSC flow itself (Supplementary Movie S2). The motion of the jump rope vortex is illustrated by the streamlines that circumscribe the vortex core in Fig.~\ref{fig:velocity}. At the centre of the vortex (pink h\"ornchen-like structure), the velocity magnitude is zero. 
As viewed in Supplementary Movie S3 and Supplementary Fig.~S2, the LSC fluid motions are in the  clockwise direction, whereas the motion of the LSC core is counter-clockwise, akin to a planetary gear.  

Looking in detail at the jump rope vortex cycle, we find at $t = 0$ that the LSC core is restricted to the horizontal midplane $z = H/2$ (Fig.~\ref{fig:velocity}\textit{B}). 
At this time, the vortex core nears the midplane sidewall, impinging on the cold downwelling flow that exists there. This splits the downwelling fluid into two branches, creating the two distinct minima in the $t=0$ midplane temperature profile (Fig.~\ref{fig:temperature}\textit{B}). The splitting motions generate horizontally divergent flows, which explain the chord probe measurements of the experimental velocity field (Fig.~\ref{fig:Doppler}\textit{D}). By necessity, on the other side of the tank, the warm upwelling flow gets collimated, creating a high pressure wave along the bottom boundary layer. We hypothesise this collimation promotes the generation of instabilities, and thereby the detachment of warm convective plumes that drive the LSC more vigorously in the lower half of the cylinder, mainly along the $\xi_{LSC}$ plane, and push the centre of the LSC upward, initiating a broadening of the warm upwelling. 

We find at $t=\tau_0/4$ (Fig.~\ref{fig:velocity}\textit{A, C}) that the vortex core no longer impinges on the midplane, but has moved to the upper half of the tank and the LSC has stretched to its longest elliptical path length. At this point, the cold downflow is still split, but the mid-plane temperature extrema are less pronounced. We find mirrored motions  at $\tau = 0$ and $\tau = \tau_0/2$, and similarly at $\tau = \tau_0/4$ and $\tau = 3\tau_0/4$, thereby producing a symmetrical jump rope cycle. 

\section*{Summary and Discussion}

We have verified that the observed jump rope behaviour is not unique to thermal convection in small $Pr$ fluids by simulating water with $Pr = 4.38$ (see SI and Fig.~S1). A coarse conditional-averaging scheme shows that the fundamental jump rope mode is detectable for $\Gamma =2$. The oscillation frequency is much lower in this fluid and the thermal and kinematic flow fields are both equally turbulent so that the sidewall temperature signals are far less pronounced (see SI). Further, such sidewall signals have also been detected in recent $Pr \simeq 12$, $\Gamma = 2$ oil experiments (Ping Wei, private communication). 

Our results significantly alter the fundamental view of large-scale circulations. Based on our combined laboratory-numerical experiments, we find that the LSC in a $\Gamma > 1$ container is not confined to a quasi-2D circulation plane perturbed by horizontal sloshing and twisting modes. Instead, we find the LSC has a dominant 3D vortex core that precesses in a jump rope-like motion in the direction opposite to that of the LSC flow field. 
%
We hypothesise that additional 3D modes exist within the LSC and that advanced techniques (e.g., dynamic mode decomposition \cite{Horn2017}, Koopman filtering \cite{Giannakis2018}) will be required to elucidate the full dynamics underlying turbulent convection in effectively unconfined natural systems.

\matmethods{
\subsection*{Laboratory Setup}
The experiments were performed with the RoMag device \citep{King2012b, King2013} using the liquid metal gallium confined in a right cylinder of aspect ratio $\Gamma=D/H=2$ with diameter $D=\SI{196.8}{mm}$ and height $H = \SI{98.4}{mm}$. The container's sidewall is made of stainless steel while the endwalls are made of copper. A non-inductively wound heater provides a heating power between \SI{6}{W} and \SI{1600}{W} at the bottom copper endwall. This heat is removed by a thermostated bath that circulates water through a double wound heat exchanger located above the top endwall. The sidewalls are wrapped by a \SI{20}{mm} layer of closed-cell foam insulation, followed by \SI{30}{mm} of Insulfrax fibrous thermal blanketing, and finally a \SI{30}{mm} layer of closed-cell foam insulation 
in order to minimise radial heat losses. 

Twenty-three experiments were conducted where the range of mean fluid temperatures varied between $\SI{35}{^\circ C} \leq T_{m} \leq \SI{47}{^\circ C}$ and the temperature drop across the fluid layer between $\SI{0.28} \leq \Delta \leq \SI{19.36}{K}$. Using the material properties for gallium \citep{Aurnou2018}, the Prandtl number ranges between $0.026 \leq Pr \leq 0.028$ and the Rayleigh number between $7.1 \times10^4 \leq Ra \leq 5.1 \times10^6$. Ultrasound Doppler velocimetry is used to measure the instantaneous velocity distribution along four different measuring lines, as shown in Fig.~\ref{fig:Doppler}. This technique is a useful tool to measure the velocities in opaque fluids such as liquid metals non-invasively \citep{Vogt2018, Vogt2013, Vogt2014}. 
The four transducers  (TR0805SS, Signalprocessing SA) capture the velocity component parallel to their ultrasound beam with a spatial resolution of about \SI{1}{mm} in beam direction and a temporal resolution of about \SI{1}{Hz}. All transducers are in direct contact with the liquid metal. They are approximately oriented in the LSC symmetry plane, except for the chord probe that is perpendicular to it. 

A total number of 29 thermocouples is used to monitor the temperatures in the experiment. Six thermistors are embedded in each of the copper endwalls \SI{2}{mm} away from the fluid layer and are used to determine the mean fluid temperature, $T_m$, and temperature drop, $\Delta$, across the fluid layer. Seven thermistors are distributed inside the fluid layer while fifteen thermocouples are placed around the perimeter outside the fluid volume at midheight $z/H=0.5$. Thirteen of those thermocouples are positioned in an array ten degrees apart and used in the experimental array in Fig.~\ref{fig:temperature}\textit{A}. The temporal resolution of the thermal measurements is \SI{10}{Hz}. Experiments are conducted until equilibration is reached, when the thermal signals vary by less than 1\% over thirty minutes. Data is then saved for between three and six thermal diffusion times. In post processing, thermocouples placed between the insulation layers provide an estimate for sidewall heat losses. Additionally, the heat losses through vertical conduction in the stainless steel sidewall are also accounted for and the top and bottom fluid temperatures are corrected to include the conduction in the copper endways.

\subsection*{Direct Numerical Simulations}
The direct numerical simulations (DNS) have been conducted with the fourth order accurate finite volume code \textsc{goldfish} \cite{Shishkina2015, Shishkina2016, Horn2017}. It numerically solves the non-dimensional Navier-Stokes equations in the Oberbeck-Boussinesq approximation augmented by the temperature equation in a cylindrical $(r, \, \phi, \, z)$ domain:
\begin{eqnarray}
  \vec{\nabla} \cdot {\vec{u}} &=& 0, \label{eq:NS1}\\
  D_{{t}} {\vec{u}} &=& Ra^{-\frac{1}{2}}
  Pr^{\frac{1}{2}} \gamma^{-\frac{3}{2}} \vec{\nabla}^2 {\vec{u}} -
  \vec{\nabla} {p} + {T} {\hat{\vec{e}}}_z, \label{eq:NS2}\\
  D_{{t}} {T} &=& Ra^{-\frac{1}{2}} Pr^{-\frac{1}{2}}
  \gamma^{-\frac{3}{2}} \vec{\nabla}^2 {T},\label{eq:NS3}
\end{eqnarray}
where $\gamma$ is the radius-to-height aspect ratio $R/H = \Gamma/2$. The radius $R$, the buoyancy velocity $(g \alpha R \Delta)^{1/2}$, the temperature difference $\Delta$ and the material properties at the mean temperature are used as the reference scales. The mechanical boundary conditions are no-slip on all solid walls, the temperature boundary conditions are isothermal for the top and bottom and perfectly insulating for the sidewalls.

The numerical resolution for the main DNS with $Pr = 0.025$, $Ra = 1.12 \times 10^6$, $\Gamma = 2$ is $N_r \times N_\phi \times N_z = 168 \times  171 \times 168$; the total run-time was 1000 free-fall time units after reaching a statistical steady-state. The obtained results were verified on a finer $280 \times 256 \times 280$ mesh. 
In addition, DNS for the same $Pr$ and $Ra$ were also carried out with a smaller aspect ratio of $\Gamma = 1$ for $t = 500 \tau_{f\!f}$. In this case, the dominant LSC motion is the commonly known combined sloshing and twisting. 
The change of behaviour of the dominant LSC mode depending on the aspect ratio was also confirmed in a high-$Pr$ fluid by means of DNS with $Pr = 4.38$ and $Ra = 10^8$ and both aspect ratios $\Gamma = 1$ and $2$, with a total run-time of  $ t = 1414\tau_{f\!f}$ and 
$ t = 612 \tau_{f\!f}$, respectively (see SI Fig.~S1). 

\subsection*{Conditional averaging} The temperature signal $T_{a\!v\!g}  = \langle T(r=R, z=H/2) \rangle_\phi$, i.e.~the sidewall temperature at midheight averaged in azimuthal direction, shows a distinct oscillation with frequency $f_0$, shown in Fig.~\ref{fig:conditional}. To extract the characteristic behaviour during one cycle, we have sampled 10 snapshots per time unit, resulting in a total of 10,000 snapshots. We then defined seven intervals based on the standard deviation $\sigma$ of $T_{a\!v\!g}$ by the boundaries $-\frac{5}{4}\sigma,\, -\frac{3}{4}\sigma,\, -\frac{1}{4}\sigma, \,\frac{1}{4}\sigma,\, \frac{3}{4}\sigma, \, \frac{5}{4}\sigma$.
For $|T_{a\!v\!g}| > \frac{5}{4} \sigma$, all snapshots in those intervals were averaged.
In the remaining intervals we additionally considered whether the signal was in a phase where the temperature increases or decreases, respectively. This was algorithmically achieved by determining if the snapshot was located between a maximum and minimum or between a maximum and minimum. However, due to the possible occurrence of several multiple local extrema during one cycle, we had to hand-select the maxima and minima as shown in Fig.~\ref{fig:conditional}. 
The result were twelve averaging intervals.

\begin{figure}[h!]
\begin{overpic}[width=\linewidth]{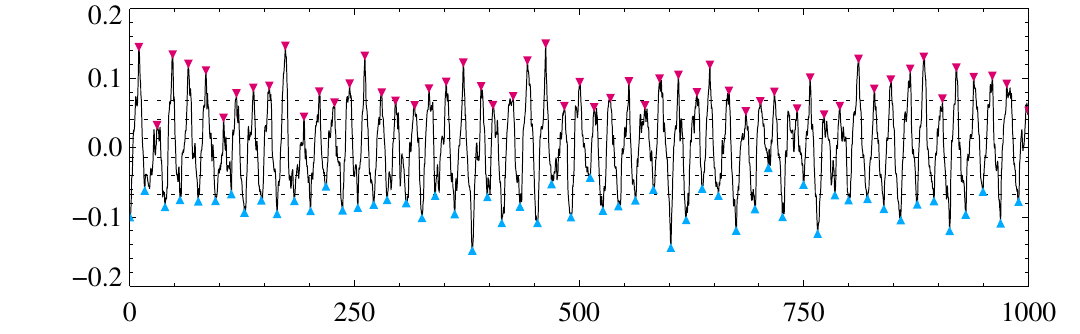}
 \put(1,16.3){\small $T_{a\!v\!g}$}
  \put(51,-2){\small $t/\tau_{f\!f}$}
\end{overpic}\\
\caption{Circumferentially-averaged temperature signal at mid-height $T_{a\!v\!g}$ from DNS with $Pr = 0.025$, $Ra = 1.12 \times 10^6$ and $\Gamma = 2$. The magenta (blue) downward (upward) triangles mark the hand-selected maxima (minima).
The horizontal dashed lines correspond to multiples of the standard deviation $\sigma$ of $T_{a\!v\!g}$, given by $\pm \frac{n}{4} \sigma, \, n \in \{1,3,5\}$, by which we separated the conditional averaging intervals.
The oscillation frequency of the signal matches $f_0$ as obtained at the centre point, shown in Fig.~\ref{fig:freq}, and is one of the most accessible identification schemes for the jump rope vortex.
\label{fig:conditional}}
\end{figure}
}
\showmatmethods{} 

\acknow{This work was supported by the NSF Geophysics Program, award no. 1547269. T.V. acknowledges the Helmholtz Association for financial support within the framework of the Helmholtz-Alliance LIMTECH and for supporting his stay in Los Angeles. S.H. acknowledges funding by the DFG under grant HO 5890/1-1 and the Leibniz-Rechenzentrum for providing computational resources on SuperMUC.}

\showacknow{} 

\bibstyle{pnas-new}
\bibliography{slosh}


\section*{Supplementary material}

# Supplementary Information for

## Jump Rope Vortex in Liquid Metal Convection

**Tobias Vogt, Susanne Horn, Alexander M. Grannan, and Jonathan M. Aurnou**

**Tobias Vogt.**
E-mail: t.vogt@hzdr.de

**This PDF file includes:**

    Supplementary text
    Figs. S1 to S2
    Captions for Movies S1 to S3

**Other supplementary materials for this manuscript include the following:**

    Movies S1 to S3

## Supporting Information Text

## Identifying the Jump Rope Vortex

The jump rope vortex leaves a characteristic imprint on several of the output parameters. It can be most easily identified using the azimuthally averaged sidewall temperature at midheight, $T_{avg}$, that is also readily available in most laboratory experiments. For $\Gamma = 2$, the signal exhibits a very regular oscillation with a clearly visible frequency $f_0$ (see Fig. 5). In contrast, the $T_{avg}$ signal in $\Gamma = 1$ does not. The thermal signal of the jump rope can also be visualised in a space-time plot (Fig. 4$A$). We show 500 free-fall time units of circumferential midplane temperatures from liquid metal and water DNS in Fig. S1$A$–$D$, as well as magnifications in $E$–$H$. The $\Gamma = 2$ data shows an accordion-like pattern in both liquid metal and water ($E, G$), although this signal is, indubitably, far easier to identify in the low $Pr$ metal. The $\Gamma = 1$ data shows a classical sloshing behaviour, i.e. the two extrema follow an in-phase zig-zag pattern ($F, H$), again with the clearer thermal signal in the thermally diffusive metal. Note that the path length of the jump rope vortex cycle is approximately $6H$, while the horizontal sloshing path length is only approximately $4R$, explaining the difference in the duration of the respective cycles for the jump roping $\Gamma = 2$ cases and the sloshing $\Gamma = 1$ cases. We have verified with DNS that both identification methods, the averaged signal $T_{avg}$ and the midplane sidewall patterns, can be used to detect jump rope or slosh modes using 8 probes uniformly placed around the circumference.

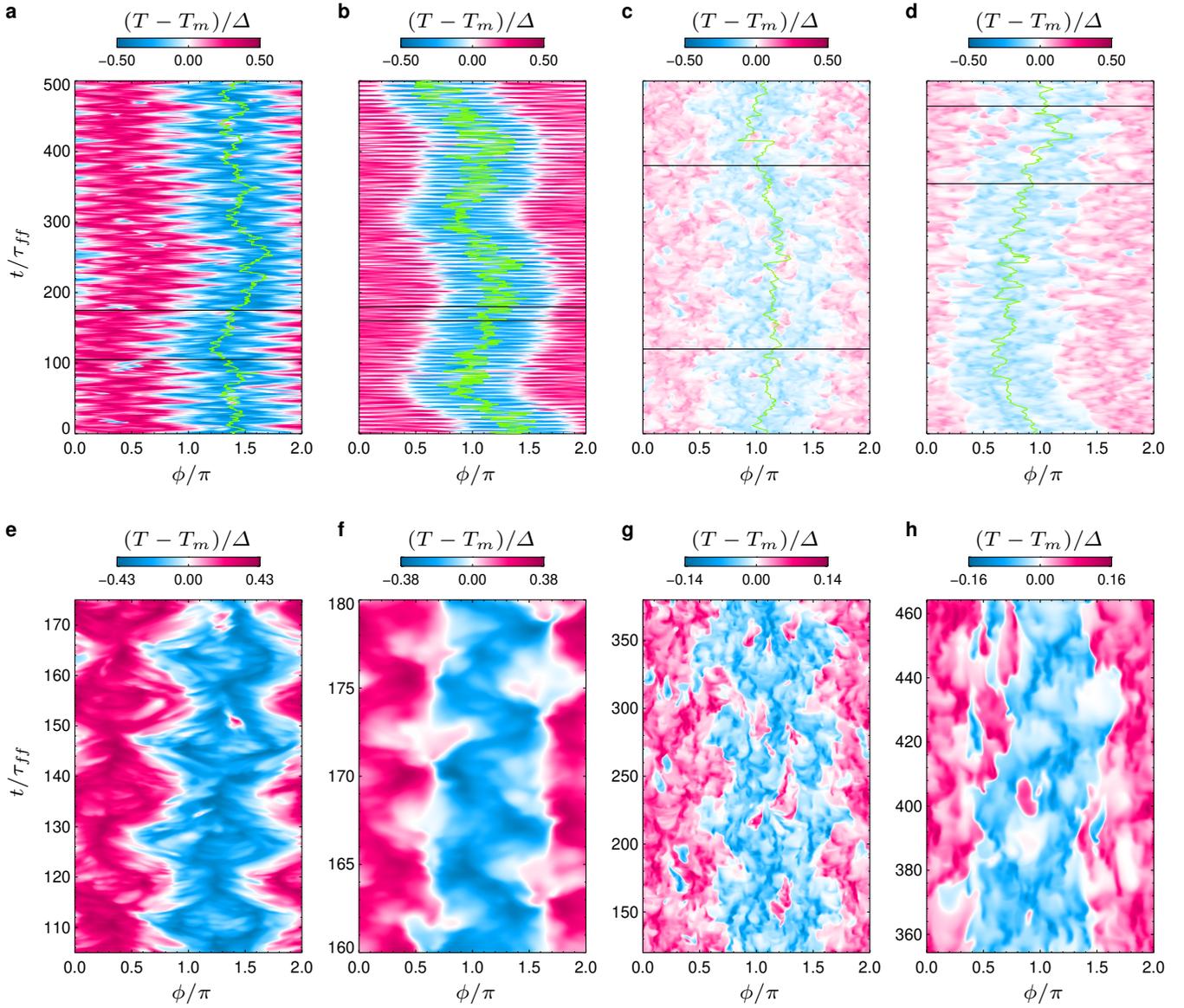

**Fig. S1.** Midplane sidewall temperature distribution for gallium and water DNS. The upper panels (A)–(D) show the sidewall thermal pattern over 500 free-fall times $t/\tau_{ff}$. The green line marks the position of the LSC $\xi_{LSC}$. The colour scale covers the entire possible temperature range, $-0.5$ to $+0.5$. The lower panels (E)–(H) are magnifications of the areas indicated by black horizontal lines in the respective panels above, showing approximately four LSC cycles and using temperature ranges restricted to the interval $[-\max|(T-T_m/\Delta|), +\max|(T-T_m/\Delta)|]$ for each case. The left-hand figures correspond to gallium, $Pr = 0.025$, $Ra = 1.12 \times 10^6$, with (A, E) $\Gamma = 2$ and (B, F) $\Gamma = 1$. The right-hand figures correspond to water, $Pr = 4.38$, $Ra = 10^8$, with (C, G) $\Gamma = 2$ and (D, H) $\Gamma = 1$.

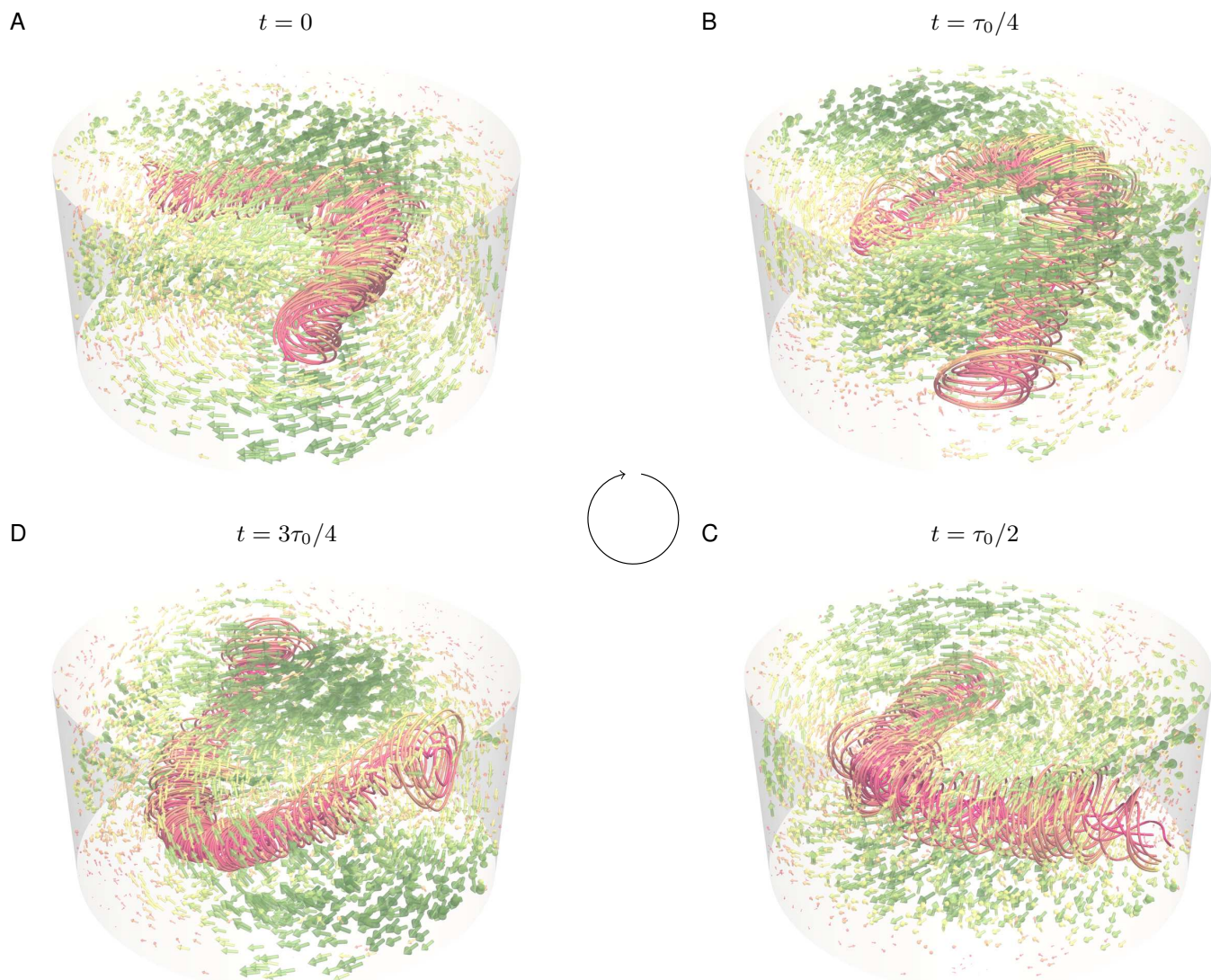

**Fig. S2.** Renderings of the LSC flow field and embedded jump rope vortex. The four panels are 3D representations of Fig. 1*B*–*D*. The vector arrows show the velocity field and are coloured and scaled in size by the magnitude $\|\vec{u}\|/u_{ff}$. The streamlines wrap around the jump rope vortex core. The colour scale is the same as in Fig. 1.

**Movie S1.** Temperature and velocity vectors of the full DNS at $Pr = 0.025$, $Ra = 1.12 \times 10^6$, $\Gamma = 2$. The time is reported in free-fall time units $\tau_{ff}$ and the temperature is scaled as $(T - T_m)/\Delta$. The left circular panel shows the temperature in the midplane. The green line marks the LSC symmetry plane $\xi_{LSC}$ and the orange line lies in the perpendicular plane $\xi_{LSC} + \pi/2$. Vertical cross-sections defined by those planes are shown in the right panels.

**Movie S2.** Conditionally-averaged temperature fields constructed using the data shown in Supplementary Movie 1, but extending out to $t/\tau_{ff} = 1000$. Note that the jump rope vortex core, shown in the upper right panel, moves opposite to the mean circulation visualised by the velocity vectors.

**Movie S3.** Jump Rope Vortex. The upper panel is colour-coded with the velocity magnitude as in Fig. 1. The lower panel shows the same streamlines and line-integral convolution (LIC) method as in the upper panel, but the colour scale corresponds to the temperature. In addition, the sidewall temperature field is displayed on the back half of the cylinder.



\end{document}